\documentclass[preprint,12pt]{elsarticle}

\usepackage{amssymb}
\usepackage{amsmath}
\usepackage{algorithm} 
\usepackage{algpseudocode} 
\usepackage{setspace}
\usepackage{lineno}
\newcommand*\patchAmsMathEnvironmentForLineno[1]{
\expandafter\let\csname old#1\expandafter\endcsname\csname #1\endcsname
\expandafter\let\csname oldend#1\expandafter\endcsname\csname end#1\endcsname
\renewenvironment{#1}
     {\linenomath\csname old#1\endcsname}
     {\csname oldend#1\endcsname\endlinenomath}}
\newcommand*\patchBothAmsMathEnvironmentsForLineno[1]{
\patchAmsMathEnvironmentForLineno{#1}
\patchAmsMathEnvironmentForLineno{#1*}}
\AtBeginDocument{
\patchBothAmsMathEnvironmentsForLineno{equation}
\patchBothAmsMathEnvironmentsForLineno{align}
\patchBothAmsMathEnvironmentsForLineno{flalign}
\patchBothAmsMathEnvironmentsForLineno{alignat}
\patchBothAmsMathEnvironmentsForLineno{gather}
\patchBothAmsMathEnvironmentsForLineno{multline}
}

\usepackage{color}

\journal{Elsevier}

\begin{document}

\begin{frontmatter}



\title{Fluid-structure interaction analysis with interface control principle} 


\author[tohoku,first,corres]{Chungil Lee} 
\author[tohoku,first,corres]{Yoshiaki Abe} 
\author[hiroshima]{Yu Kawano} 
\author[tohoku2,tohoku]{Tomoki Yamazaki} 
\affiliation[tohoku]{organization={Tohoku University, Institute of Fluid Science},
            addressline={2-1-1 Katahira, Aoba-ku}, 
            city={Sendai},
            postcode={9808577}, 
            state={Miyagi},
            country={Japan}}
\affiliation[tohoku2]{organization={Tohoku University, Department of Aerospace Engineering},
            addressline={6-6-01 Aramaki-Aza-Aoba, Aoba-ku}, 
            city={Sendai},
            postcode={9808579}, 
            state={Miyagi},
            country={Japan}}
\affiliation[hiroshima]{organization={Hiroshima University, Graduate School of Advanced Science and Engineering},
            addressline={1-4-1 Kagamiyama}, 
            city={Higashi-Hiroshima},
            postcode={739-8527}, 
            state={Hiroshima},
            country={Japan}}
\affiliation[corres]{Corresponding author} 
\affiliation[first]{Chungil Lee and Yoshiaki Abe equally contributed to this paper.}

\begin{abstract}
An interface control principle is proposed for unsteady fluid-structure interaction (FSI) analyses. This principle introduces a method of explicitly controlling the interface motion in the temporal direction to minimize the residual force on the interface, which is defined as the discrepancy between the fluid and structural forces. The interface model is constructed using a data-driven approach that involves sparse identification of nonlinear dynamics with control to evaluate the residual force. The interface model is subsequently subjected to control theory in order to minimize the residual force. Following the convergence of the residual force, the interface state is controlled to be that of the original unsteady FSI system. The fluid and structural simulations can be conducted independently without communication, as the interface state information is predetermined as inputs for each system. The proposed method is applied to the vortex-induced vibration (VIV) of a cylinder at a Reynolds number of 150 under several reduced velocity conditions corresponding to the lock-in regime with limit-cycle oscillations. The results demonstrate that the residual force is sufficiently minimized in time, and when the residual force is close to zero, the predicted fluid force and structural displacement of the VIV show good agreement with the reference FSI simulation.
\end{abstract}


\end{frontmatter}


\setstretch{1.1}
\section{Introduction}
Multi-physics problems such as fluid-structure interaction (FSI), constitute an essential field in both fundamental and pragmatic engineering topics~\citep{Date2022}. Recent advancements in simulation technology for fluid and structural analysis facilitate high-fidelity simulations, including large-eddy simulation (LES) in fluid dynamics~\citep{Abe2023} and geometrical or material nonlinear finite element analysis (FEM) in solid mechanics~\citep{Liu2025}. Such high-fidelity and essentially unsteady simulations can provide accurate solutions, but they come with significant computational costs due to the large number of spatial degrees of freedom and the extended time periods required. Therefore, it is essential to develop methods for FSI problems that reduce computational costs while maintaining the ability to handle the large degrees of freedom needed to analyze real-world problems.

A coupling strategies for FSI problems are usually classified into two approaches: monolithic and partitioned approaches. The monolithic approach integrates fluid and structure systems by simultaneously solving their governing equations \citep{heil2004efficient,bathe2004finite,ishihara2005monolithic,tezduyar2006solution}. This approach can provide highly accurate solutions to FSI problems due to its ability to exactly meet the coupling conditions between the fluid and structural domains.
In the meantime, the stable formulation of numerical algorithms frequently presents mathematical challenges and constraints on the application of high-fidelity solvers.
The partitioned method solves the fluid and solid systems separately, allowing high-fidelity simulations to be applied to each physical domain independently \citep{felippa2001partitioned}. 
In the partitioned method, the one-way coupling (or loosely-coupling) approach transfers quantities between systems at each or specified time step without considering exact coupling conditions \citep{farhat2000two}. This method is computationally efficient; however, it often leads to unstable and inaccurate simulations in systems characterized by strong interactions \citep{causin2005added}.
Another approach is the two-way coupling (or strong coupling), which involves iteratively solving both systems and exchanging interface quantities until the coupling criterion is satisfied.  This method generally enables more stable and accurate simulations; however, the computational expenses related to its iterative calculations are significantly higher than those of the one-way coupling approach \citep{degroote2008stability,kuttler2008fixed,bogaers2014quasi}.

In both one-way and two-way couplings, partitioned approaches are preferred over monolithic methods due to their non-intrusive formulation, which simplifies the integration of existing simulation codes. Meanwhile, transferring quantities on the interface necessitates interaction between fluid and structural systems, frequently resulting in inefficient code implementation and redundant communication across multiple computational threads or nodes in large-scale simulations. Thus far, this bottleneck has been unavoidable as a result of an inherent characteristic of the iterative convergence approach for partitioned coupling strategies.

On the other hand, a reduced-order model (ROM) has also received attention as a solution to this problem by transforming higher-dimensional data into lower-dimensional data while preserving the primary features of the original system. The proper orthogonal decomposition (POD) introduced by \citet{berkooz1993proper} is one of the most widely used methods for ROM and is utilized as an effective tool for dimensionality reduction. In FSI problems, POD extracts low-dimensional data from high-fidelity data, and ROM are then constructed by projecting the governing equations onto the low-order POD basis \citep{lieu2006reduced,barone2009reduced,liberge2010reduced,kalashnikova2013stable,longatte2018parametric}. These projection methods, such as POD-Galerkin projection, are typically classified as model-based approaches and can effectively yield the FSI solution. Meanwhile, access to the source code that describes the physical system is essential for constructing the projection step, due to its dependency on the governing equations.

To mitigate the aforementioned disadvantage, several non-intrusive ROM methods have been proposed \citep{xiao2016non,shinde2019galerkin,azzeddine2024non}. These approaches are purely data-driven methods for ROM in FSI problems, eliminating the need for the governing equations. In non-intrusive ROM methods, POD is used for dimensionality reduction similarly to intrusive ROM methods, but instead of projecting the higher-dimensional representation, POD modes and corresponding temporal coefficients are employed to estimate high-fidelity data. Recently, non-intrusive ROM approaches using deep neural network (DNN) have been also developed \citep{cheng2021deep,zhang2022data,han2022deep,liu2024novel,lee2024parametric}. \citet{zhang2022data}  employed a convolutional variational autoencoder model to map the high-dimensional dynamics of the three-dimensional vortex-induced vibration (VIV) of a flexible cylinder onto a low-dimensional latent space. They were able to predict the motions and flow fields of VIV at future time steps with high accuracy by using a governing equation of the reduced latent vector, obtained through the sparse identification of nonlinear dynamics (SINDY) algorithm. \citet{han2022deep} and \citet{liu2024novel} proposed the DNN-based ROM to replace computationally expensive CFD simulations and predicted the flow field at the next time step using the flow fields and structural motions at the current step. The structural motions at the next time step are then determined by solving the equation of structural motion based on the predicted flow field. The flow field and structural motions at future time steps can be obtained by performing this process iteratively.

Although these data-driven approaches have been well demonstrated to recover the trained dataset, they have not been sufficiently validated for the accuracy on out-of-sample data, raising concerns about their generalizability \citep{han2022deep,liu2024novel}. This is because these models may not inherently respect the underlying physics, leading to non-physical results when applied to conditions outside of the training dataset. Therefore, this study aims to develop an integrated method that combines the strengths of data-driven and model-based methods, while also accounting for the system separability, a significant benefit of partitioned methods. For this purpose, the model-based approach is partly employed to achieve a physically consistent interface condition, while iterative calculations are bypassed through the application of control theory to the data-driven system that represents the interface status.

In this study, we propose a residual force minimization approach based on an interface control principle.
This approach facilitates the coupling of fluid and structural domains without the need for an iterative process, by employing control theory in developing an interface state that minimizes residual force over time.
The residual force is defined as the discrepancy between the fluid and structural forces at the interface. A dynamical system is developed to estimate the residual force through reduced-order modeling, with the interface state controlled to minimize the residual force.
The proposed method fundamentally differs from the partitioned method and the aforementioned data-driven ROM techniques, offering a novel approach to mitigate implementation costs for multi-physics simulation codes while also decreasing computational expenses. 
The rest of this article is organized as follows. In Section~2, residual force minimization approach based on interface control principle is presented. Then, Section~3 describes its application to the VIV problem. Conclusions are finally provided in Section~4. 


\begin{figure}
  \begin{center}
\includegraphics[width=137mm]{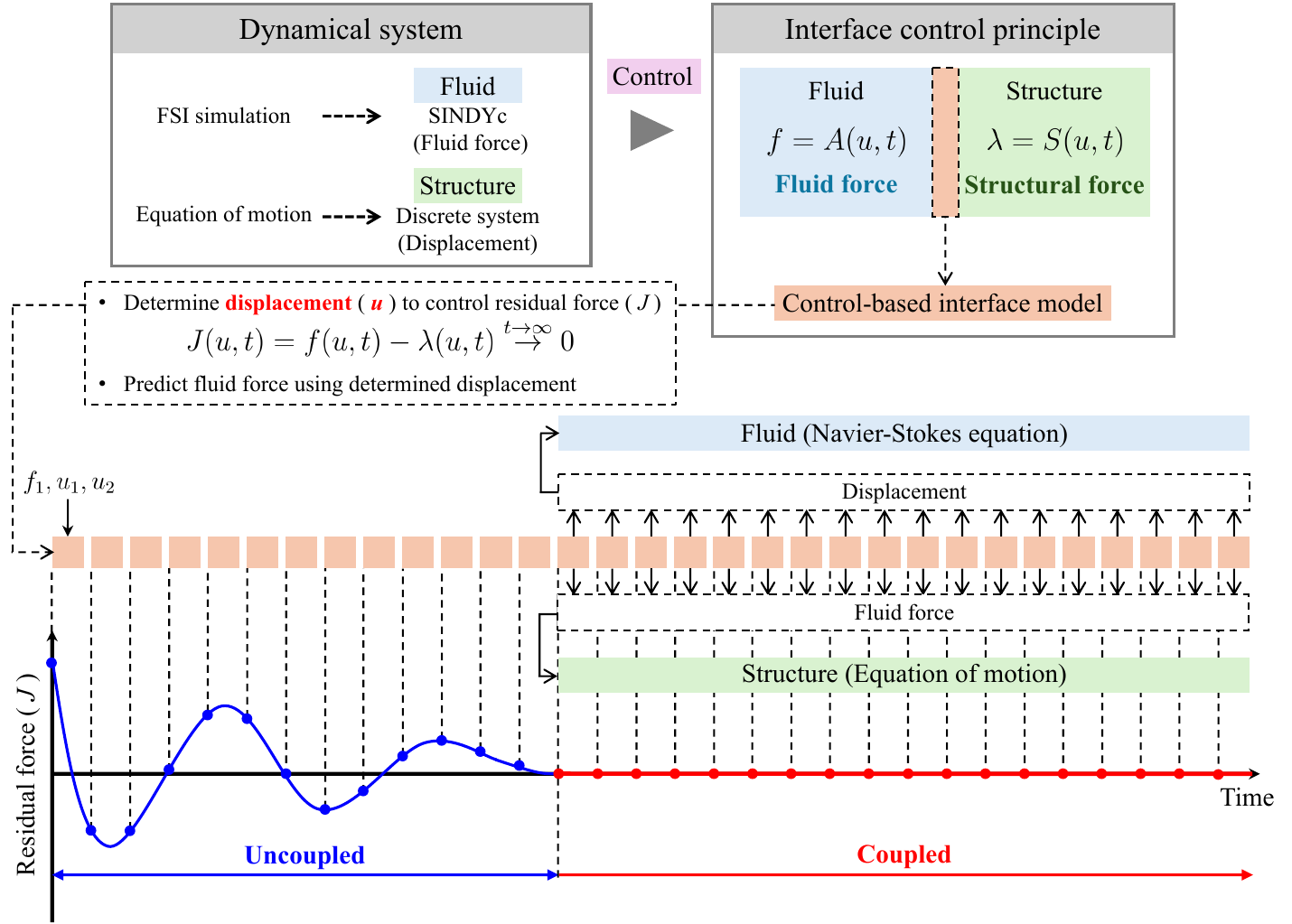}
\caption{Schematic of the interface control principle for the residual force minimization.}
\label{fig:Schematic of the proposed method}
\end{center}
\end{figure}

\section{\label{sec:Residual force minimization}Interface control principle for residual force minimization}
A schematic of the interface control principle for the residual force minimization method is shown in Fig.~\ref{fig:Schematic of the proposed method}. 
First, dynamical systems that predict the future states of the physical quantities in each domain are constructed. 
This study focuses on the VIV problem for demonstration, and thus the dynamical system for the structural displacement is represented by a single equation of motion.
On the other hand, the dynamical system for the fluid force is modeled by the SINDY with control (SINDYc) which is nonlinear system identification \citep{brunton2016sparse,kaiser2018sparse}. In construction of the SINDYc, the fluid force and structural displacement obtained by the reference FSI simulation are utilized.

A control-based interface model is then constructed to explicitly minimize the residual force in time by designing the structural displacement as a control input.
In the present study, the fluid force is simultaneously predicted through the designed structural displacement and the dynamical system for the fluid force.
The residual force in this context denotes the discrepancy between the fluid and structural forces at the interface. Consequently, the presence of the residual force suggests that the interface is not in a coupled state between the dynamical systems of fluid and structure.
Then, the structural displacement in the coupled state is recovered by the control input after the residual force has vanished, and it can be used as inputs for the fluid simulation.
As such, a fundamental concept of the interface control principle is the implementation of the control-based interface model that facilitates explicit regulation of the interface into the coupled state of dynamical systems.

This approach is different from the existing partitioned approach in that it does not necessitate direct communication between the fluid and structure systems. Instead, the interface model is positioned in between to establish the appropriate coupled state by eliminating the residual force.
The following subsections provide detailed descriptions of the discrete dynamical systems of the physical quantities, the control principle, and the control-based interface model.

\subsection{\label{subsec:linear system}Discrete dynamical system for structure}
The dynamical system for structural displacement is assumed to be represented by the linear equation of motion in this study. The discrete system for the structural displacement is expressed as:
\begin{align}
\label{eq:structural system}
    \textbf{u}_{k+1} & = \textbf{A}_s\textbf{u}_{k}+\textbf{B}_s\textbf{x}_{k},
\end{align}
where $\textbf{u}_k\in \mathbb{R}^{p_u}$ and $\textbf{x}_k\in \mathbb{R}^{p_x}$ are vectors for the structural displacement and the fluid force at the time step $k$, respectively. 
$p_{u}$ and $p_{x}$ denote the number of variables to represent the structural displacement and fluid force, respectively, in the discrete system.
The matrices $\textbf{A}_s\in \mathbb{R}^{p_u\times p_u}$ and $\textbf{B}_s\in \mathbb{R}^{p_u\times p_x}$ are obtained by converting a continuous system of the equation of structural motion into a discrete system.
It should be noted that a data-driven system identification, which is described in the next subsection for fluid, can be also employed to consider the more general structural problem.

\subsection{\label{subsec:linear system}Discrete dynamical system for fluid}
This study employs a data-driven approach to establish the dynamical system estimating fluid force.
The SINDYc method \citep{brunton2016sparse} is utilized as a sparsity-promoting technique for identifying nonlinear dynamical systems with inputs $\textbf{u}(t)$. Generally, the dynamical system with $\textbf{u}(t)$ can be represented by
\begin{eqnarray}
    \frac{{\rm d}}{{\rm d}t}\textbf{x}(t)=\textbf{f}(\textbf{x}(t),\textbf{u}(t)),
    \label{eq:system}
\end{eqnarray}
where $\textbf{x}(t)\in \mathbb{R}^{p_x}$ and $\textbf{u}(t)\in \mathbb{R}^{p_u}$ represent the state and control input of the system at time $t$, respectively, and $\textbf{f}\in \mathbb{R}^{p_x}$ denotes a sparse function that defines the system dynamics. To identify the function $\textbf{f}$ using the SINDYc algorithm, the time-series data of the state and input variables are first organized into two matrices:
\begin{align}
\textbf{X}=\begin{bmatrix}
\textbf{x}^{\mathsf{T}}(t_1) \\
\textbf{x}^{\mathsf{T}}(t_2) \\
\vdots \\
\textbf{x}^{\mathsf{T}}(t_m)
\end{bmatrix}=\begin{bmatrix}
x_1(t_1) & x_2(t_1) & \cdots  & x_{p_x}(t_1) \\
x_1(t_2) & x_2(t_2) & \cdots  & x_{p_x}(t_2) \\
\vdots  &  \vdots &  \ddots &  \vdots\\
 x_1(t_m) & x_2(t_m) & \cdots  & x_{p_x}(t_m) \\
\end{bmatrix}\in \mathbb{R}^{m\times p_x},\\
 \textbf{U}=\begin{bmatrix}
\textbf{u}^{\mathsf{T}}(t_1) \\
\textbf{u}^{\mathsf{T}}(t_2) \\
\vdots \\
\textbf{u}^{\mathsf{T}}(t_m)
\end{bmatrix}=\begin{bmatrix}
u_1(t_1) & u_2(t_1) & \cdots  & u_{p_u}(t_1) \\
u_1(t_2) & u_2(t_2) & \cdots  & u_{p_u}(t_2) \\
\vdots  &  \vdots &  \ddots &  \vdots\\
 u_1(t_m) & u_2(t_m) & \cdots  & u_{p_u}(t_m) \\
\end{bmatrix}\in \mathbb{R}^{m\times p_u},
\end{align}
where $m$ is the number of snapshots, and the notation $\circ^{\mathsf{T}}$ indicates the transpose. The library of candidate nonlinear functions $\Theta$ is then constructed using data matrices $\textbf{X}$ and $\textbf{U}$:
\begin{align}
\Theta(\textbf{X},\textbf{U}) =\begin{bmatrix}
 |& | & | & | & | &    \\
 \textbf{1}& \textbf{X} &\textbf{U}  & \left ( \textbf{X}\otimes\textbf{X} \right ) & \left ( \textbf{X}\otimes\textbf{U} \right )  & \cdots  \\
 |& | & | & | & | &   \\
\end{bmatrix}, 
\end{align}
where $\textbf{X}\otimes\textbf{U}$ denotes the vector comprising all pairwise products between the components $\textbf{X}$ and \textbf{U}. 
The design of a suitable library of candidate terms is crucial to the performance of the SINDYc algorithm. Finally, the system given by Eq.~\eqref{eq:system} can be written as:
\begin{eqnarray}
\dot{\textbf{X}}=\Theta(\textbf{X},\textbf{U})\Xi,
\end{eqnarray}
where
\begin{eqnarray}
   \dot{\textbf{X}}=\begin{bmatrix}
\dot{\textbf{x}}^{\mathsf{T}}(t_1) \\
\dot{\textbf{x}}^{\mathsf{T}}(t_2) \\
\vdots \\
\dot{\textbf{x}}^{\mathsf{T}}(t_m)
\end{bmatrix}=\begin{bmatrix}
\dot{x}_1(t_1) & \dot{x}_2(t_1) & \cdots  & \dot{x}_{p_x}(t_1) \\
\dot{x}_1(t_2) & \dot{x}_2(t_2) & \cdots  & \dot{x}_{p_x}(t_2) \\
\vdots  &  \vdots &  \ddots &  \vdots\\
 \dot{x}_1(t_m) & \dot{x}_2(t_m) & \cdots  & \dot{x}_{p_x}(t_m) \\
\end{bmatrix} 
\end{eqnarray}
The notation $\dot{\circ}$ represents the differentiation in time, and the matrix $\dot{\textbf{X}}$ is obtained by the second-order central-difference scheme in this study. The coefficient matrix $\Xi$ is composed of sparse column vectors $\xi$, each of which determines the active terms from $\Theta(\textbf{X},\textbf{U})$ in the dynamics of $\textbf{f}$. These vectors are obtained using a convex $l_1$ regularized sparse regression:
\begin{eqnarray}
    \xi_{i}=\text{argmin}_{\xi_{i}} \left\| \dot{\textbf{X}}_i-\Theta(\textbf{X},\textbf{U})\xi_{i}\right\|_{2}+\alpha\left\| \xi_{i}\right\|_{1},
\end{eqnarray}
where $\dot{\textbf{X}}_i$ and $\xi_{i}$ denote the $i$-th column of $\dot{\textbf{X}}$ and $\Xi$, respectively. 
The $\left\|\circ\right\|_1$ term promotes sparsity in the vector $\xi_{i}$.  The parameter $\alpha$ represents the weight of the constraint term and determines the sparsity of the coefficient vector $\xi$. In the present work, a sequentially thresholded least squares algorithm is used to solve this sparsity-promoting regression problem. When the matrix $\Xi$ has been determined, the nonlinear dynamical systems with inputs at the time step $k$ can be expressed as:
\begin{eqnarray}
\dot{\textbf{x}}^{\mathsf{T}}_k=\Theta (\textbf{x}_k^{\mathsf{T}},\textbf{u}_k^{\mathsf{T}})\Xi. 
\label{eq:SINDYc_decrete}
\end{eqnarray}
Therefore, the dynamical system for the fluid force can be written from Eq.~\eqref{eq:SINDYc_decrete}, as follows:
\begin{eqnarray}
    \textbf{x}_{k+1}=\textbf{x}_{k}+\Delta t\Xi^{\mathsf{T}}\Theta ^{\mathsf{T}}(\textbf{x}^{\mathsf{T}}_{k},\textbf{u}^{\mathsf{T}}_{k}),
    \label{eq:fluid system}
\end{eqnarray}
where $\Delta t$ is the time interval. 

\subsection{\label{subsec:Control principle}Control principle}
The control-based interface model minimizes the residual force by designing structural displacement. 
Unlike the conventional fixed-time iteration employed in partitioned methods, our approach seeks to identify the evolution of the interface to minimize the residual force over time, as illustrated in Fig.~\ref{fig:Schematic of the proposed method}.
The residual force $\textbf{J}_{k}$ in the discrete-time system can thus be defined through the discrete system for structure, as follows:
\begin{align}
    \textbf{J}_{k} & = \textbf{u}_{k+1}-\textbf{A}_s\textbf{u}_{k}-\textbf{B}_s\textbf{x}_{k}.
\label{eq:residual force} 
\end{align}
Here, $\textbf{u}_{k+1}-\textbf{A}_s\textbf{u}_{k}$ and $\textbf{B}_s\textbf{x}_{k}$ correspond to the structural and fluid force terms, respectively. In the presence of residual force, the structural system can be described by Eq.~\eqref{eq:residual force}, rather than Eq.~\eqref{eq:structural system}. We then design the following discrete-linear system of the residual force:
\begin{align}
    \textbf{J}_{k+1} & = \textbf{C}\textbf{J}_{k}.
    \label{eq:residual force system} 
\end{align}
The stability of the system in Eq.~\eqref{eq:residual force system} is determined by the eigenvalues of the matrix $\textbf{C}\in \mathbb{R}^{p_u\times p_u}$. In other words, the evolution of the residual force is explicitly controlled by these eigenvalues. If all eigenvalues of the matrix $\textbf{C}$ are inside the unity circle in the complex plane, the residual force decays over time. To ensure this behavior, the matrix $\textbf{C}$ is defined as a scalar matrix, $\textbf{C}=\lambda\textbf{I}$, where $\lambda$ is a scalar constant whose absolute value is less than one;
$\textbf{I}\in \mathbb{R}^{p_u\times p_u}$ denotes the identity matrix. Hereafter, the value $\lambda$ is referred to as the {\it control parameter}.

\subsection{\label{subsec:interface model}Control-based interface model}
\begin{figure}
  \begin{center}
\includegraphics[width=120mm]{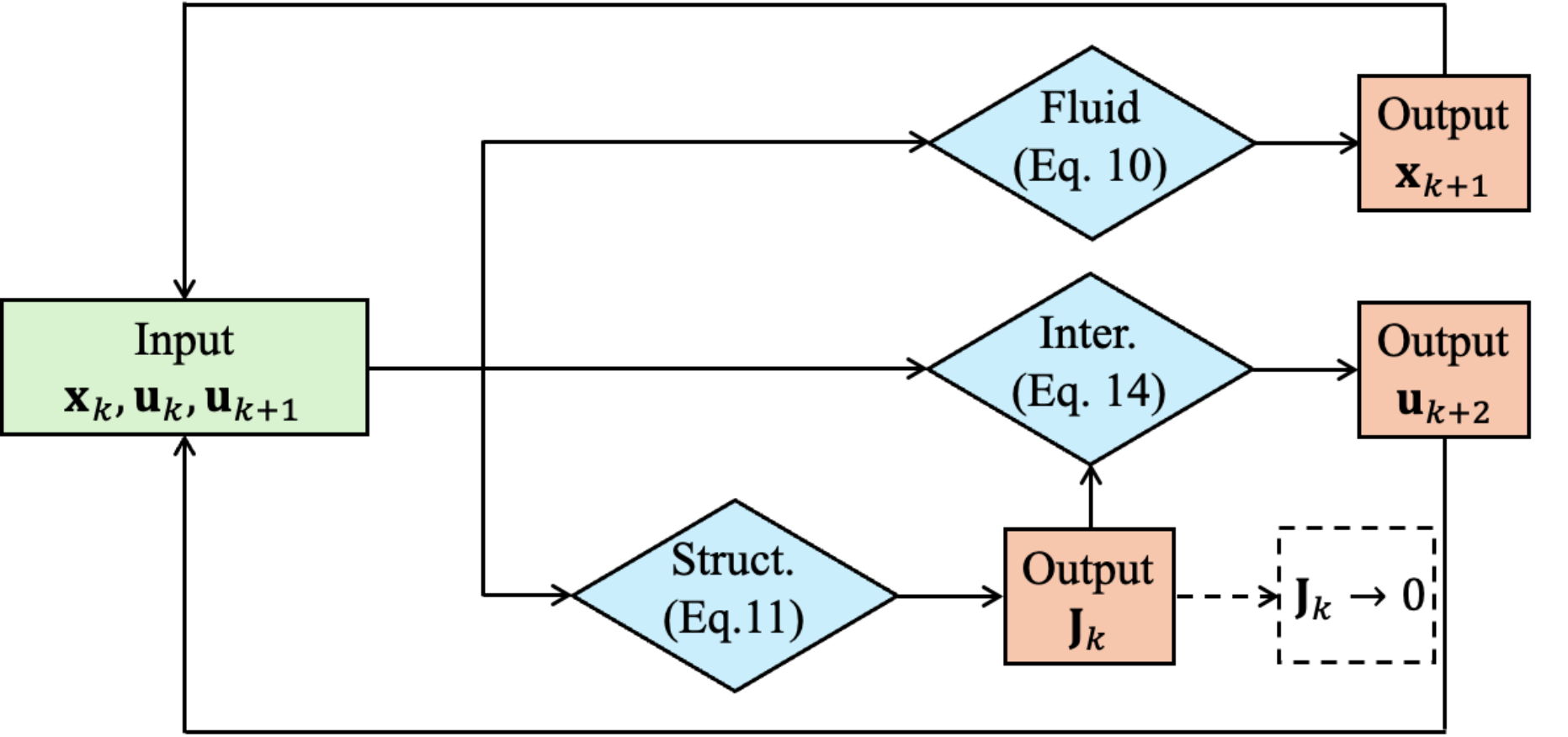}
\caption{Flow chart of the proposed method for minimizing the residual force and predicting the physical quantities.}
\label{fig:Flow chart}
\end{center}
\end{figure}

The control-based interface model is constructed based on the discrete-linear system of the residual force given by Eq.~\eqref{eq:residual force system}. First, the residual force at time step $k+1$ is expressed by Eq.~\eqref{eq:residual force} and can be rewritten by substituting the discrete systems for the physical quantities, given in Eqs.~\eqref{eq:fluid system} and \eqref{eq:residual force} as follows:
\begin{align}
 \textbf{J}_{k+1} &= \textbf{u}_{k+2}-\textbf{A}_s\textbf{u}_{k+1}-\textbf{B}_s\textbf{x}_{k+1} \nonumber\\
 & =  \textbf{u}_{k+2}-\textbf{A}_s^2\textbf{u}_{k}-(\textbf{I} + \textbf{A}_s)\textbf{B}_s\textbf{x}_{k}-\Delta t \textbf{B}_{s}\Xi^{\mathsf{T}}\Theta ^{\mathsf{T}}(\textbf{x}^{\mathsf{T}}_{k},\textbf{u}^{\mathsf{T}}_{k})-\textbf{A}_s\textbf{J}_k.
 \label{eq:next residual force system}
\end{align}
Then, by substituting Eq.~\eqref{eq:next residual force system} into Eq.~\eqref{eq:residual force system}, the following system is obtained:
\begin{eqnarray}
    \textbf{u}_{k+2}=\textbf{A}_s^2\textbf{u}_{k} + (\textbf{I} + \textbf{A}_s)\textbf{B}_s\textbf{x}_{k} +\Delta t \textbf{B}_{s}\Xi^{\mathsf{T}}\Theta ^{\mathsf{T}}(\textbf{x}^{\mathsf{T}}_{k},\textbf{u}^{\mathsf{T}}_{k})+(\textbf{A}_s+\textbf{C})\textbf{J}_{k}.
    \label{eq:control based ver1}
\end{eqnarray}
This system decreases the residual force by controlling structural displacement at each time step, as illustrated in Fig.~\ref{fig:Flow chart}. The input data comprises the fluid force and structural displacement at time step $k$, along with the structural displacement at time step $k+1$, which is used to calculate the residual force at time step $k$. The fluid force and structural displacement at the subsequent time step (i.e., output data) are then predicted using the discrete systems for the physical quantities and the control-based interface model. This closed-loop process iteratively estimates fluid force and structural displacement while progressively minimizing the residual force.

\begin{algorithm}
\caption{Control-based interface model}\label{alg:interface model}
\textbf{Input}: Initial conditions $\textbf{x}_1$, $\textbf{u}_1$, $\textbf{u}_{2}$
\begin{algorithmic}[1]
\While{$k=1,\cdots,m$}
\State
\noindent $
\textbf{J}_{k} = \textbf{u}_{k+1}-\textbf{A}_s\textbf{u}_{k}-\textbf{B}_s\textbf{x}_{k}$\Comment{Eq.~\eqref{eq:residual force}}
\State $\textbf{u}_{k+2}=\textbf{A}_s^2\textbf{u}_{k} + (\textbf{I} + \textbf{A}_s)\textbf{B}_s\textbf{x}_{k} +\Delta t \textbf{B}_{s}\Xi^{\mathsf{T}}\Theta ^{\mathsf{T}}(\textbf{x}^{\mathsf{T}}_{k},\textbf{u}^{\mathsf{T}}_{k})+(\textbf{A}_s+\textbf{C})\textbf{J}_{k}$\\ \Comment{Eq.~\eqref{eq:control based ver1}}
\State $\textbf{x}_{k+1}=\textbf{x}_{k}+\Delta t\Xi^{\mathsf{T}}\Theta ^{\mathsf{T}}(\textbf{x}^{\mathsf{T}}_{k},\textbf{u}^{\mathsf{T}}_{k})$ \Comment{Eq.~\eqref{eq:fluid system}}
\EndWhile
\end{algorithmic}
\end{algorithm}

\section{Demonstration}
In this section, we examine the residual minimization approach based on the interface control principle through its application to the VIV of the cylinder. 

\subsection{Problem settings for vortex-induced vibration of cylinder}
\begin{figure}
  \begin{center}
\includegraphics[width=60mm]{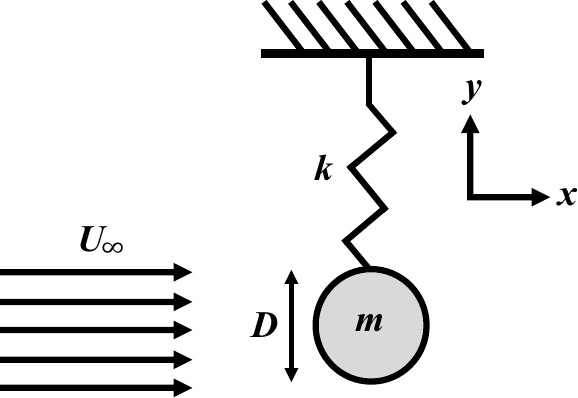}
\caption{Schematic of VIV for two-dimensional cylinder.}
\label{fig:Schematic of VIV}
\end{center}
\end{figure}

A two-dimensional cylinder is placed in a freestream, as shown in Fig.~\ref{fig:Schematic of VIV}. The cylinder is supported by a spring which allows vertical oscillation. The Reynolds number defined by $U_{\infty}D/ \nu$ is set to be 150, where $U_\infty$, $D$, and $\nu$ are the freestream velocity, the cylinder diameter, and the kinematic viscosity, respectively~\cite{ahn2006strongly,borazjani2008curvilinear}.

In the FSI analysis for the reference, the two-dimensional compressible Navier-Stokes equations were coupled with an equation of motion for the cylinder, which were solved by an in-house finite-difference solver LANS3D~\citep{fujii1989high,abe2014geometric,Abe2023}.
The inflow Mach number calculated by $M_{\infty}=U_{\infty}/a_{\infty}$ was set to be 0.2, where $a_{\infty}$ is the speed of sound. The Mach number is sufficiently low to avoid the influence of compressibility. The spatial derivatives were evaluated by a sixth-order compact finite-difference scheme, and the tenth-order filtering with the filter coefficient of 0.495 was used to suppress a numerical oscillation. A second-order implicit time scheme was employed for advancing time. An O-type moving mesh was utilized, with a grid of 201 $\times$ 201 points distributed along the circumference and wall height directions. The mesh was moved without deformation according to the equation of motion for the cylinder. The computation was performed for 200,000 steps, and the time step size normalized by a speed of sound and cylinder diameter, was set to be 0.005.

The equation of motion for the two-dimensional cylinder is expressed in a non-dimensional variable as follows:
\begin{align}
\label{eq:structural VIV}
    \ddot{y}^{*}+\left ( \frac{4\pi\xi}{U^{*}}\right )\dot{y}^{*}+\left ( \frac{2\pi}{U^{*}}\right )^{2}M_{\infty}^{2}{y}^{*}=\frac{M_{\infty}^{2}}{2m^{*}}C_{L},
\end{align}
where $\xi$ and $C_{L}$ are the non-dimensional damping coefficient and lift coefficient, respectively. The equation~\eqref{eq:structural VIV} is normalized by using the following relations:
\begin{eqnarray}
     y^{*}=\frac{y}{D}, \quad t^{*}=\frac{t}{D/a_{\infty}}, \quad m^{*}=\frac{m}{\rho_{\infty}D^{2}}, \quad
     U^{*}=\frac{U_{\infty}}{fD}, \quad C_{L}=\frac{2\lambda_{y}}{\rho_{\infty}U_{\infty}^{2}D},
\end{eqnarray}
where $y^{*}$, $t^{*}$, and $m^{*}$ are the non-dimensional displacement, time, mass of the cylinder. The value $U^{*}$ is the reduced velocity, $f$ is the natural frequency of the cylinder, $\rho_{\infty}$ is the density of the fluid, and $\lambda_{y}$ is the fluid force in the direction of $y$. In this study, the damping term is neglected ($\xi=0$), as shown in Fig.~\ref{fig:Schematic of VIV}.
Defining the structural and fluid force state vectors as $\textbf{u}_{k} =
\begin{bmatrix}
     y^{*}_{k} & \dot{y}^{*}_{k}
\end{bmatrix}^{\mathsf{T}}$ and $\textbf{x}_{k} =
\begin{bmatrix}
     C_{L,k} & \dot{C}_{L,k}
\end{bmatrix}^{\mathsf{T}}$, respectively, the matrices $\textbf{A}_s$ and $\textbf{B}_s$ in Eq.~\eqref{eq:structural system} can be derived from the equation of motion as follows:
\begin{align}
\textbf{A}_s&=\left ( \textbf{I}_2+\Delta t \begin{bmatrix}
0 &  1\\
 -\left ( \frac{2\pi}{U^{*}}\right )^{2}M_{\infty}^{2}&  0\\
\end{bmatrix} \right ), \\
\textbf{B}_s&=\Delta t\begin{bmatrix}
0 & 0  \\
\frac{M_{\infty}^{2}}{2m^{*}} & 0 \\
\end{bmatrix},
\end{align}
where $\textbf{I}_2 $ represents the $2 \times 2$ identical matrix.

\begin{figure}
  \begin{center}
\includegraphics[width=100mm]{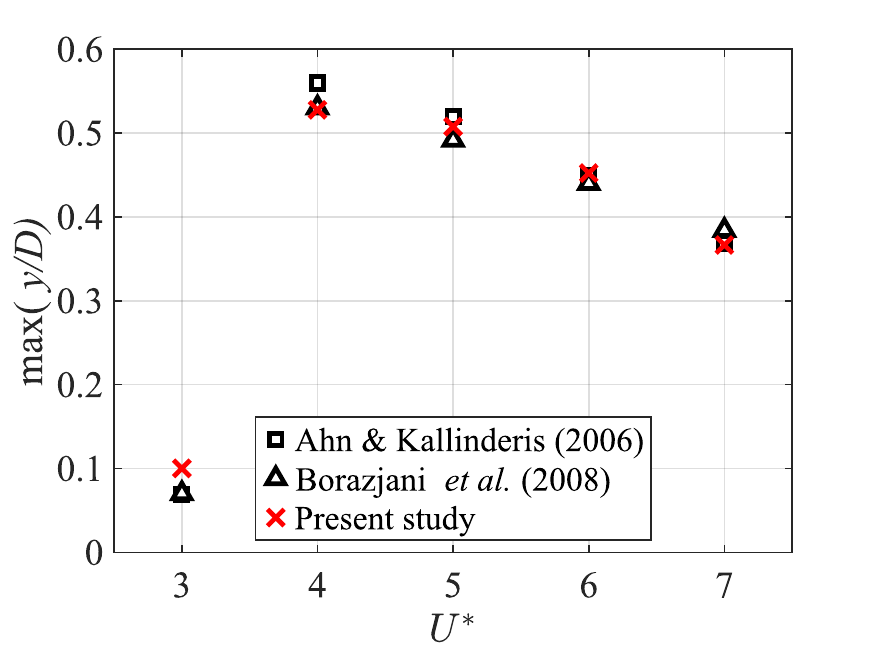}
\caption{Comparisons of the maximum cylinder displacements for each $U^{*}$ between the present work and previous studies \citep{ahn2006strongly,borazjani2008curvilinear} }
\label{fig:lock-in}
\end{center}
\end{figure}

In the VIV, the reduced velocity $U^*$ characterizes the resonance oscillation of the system, commonly known as the lock-in phenomenon. Figure~\ref{fig:lock-in} shows the simulation results for the maximum displacements of the cylinder depending on the reduced velocity. The maximum displacements increase significantly in the range of $4 \leq U^{*} \leq 7$, whereas they decrease outside this range. Therefore, the lock-in phenomenon is observed in $4 \leq U^{*} \leq 7$. These results are validated by comparison with the previous results of \citet{ahn2006strongly} and \citet{borazjani2008curvilinear}. This study focused on the VIV at reduced velocities within the lock-in regime ($U^{*}=4,5,6,7$) to demonstrate the proposed method.

\subsection{\label{subsub:VIV system validation}Validation of discrete fluid system}
\begin{figure}
  \begin{center}
\includegraphics[width=135mm]{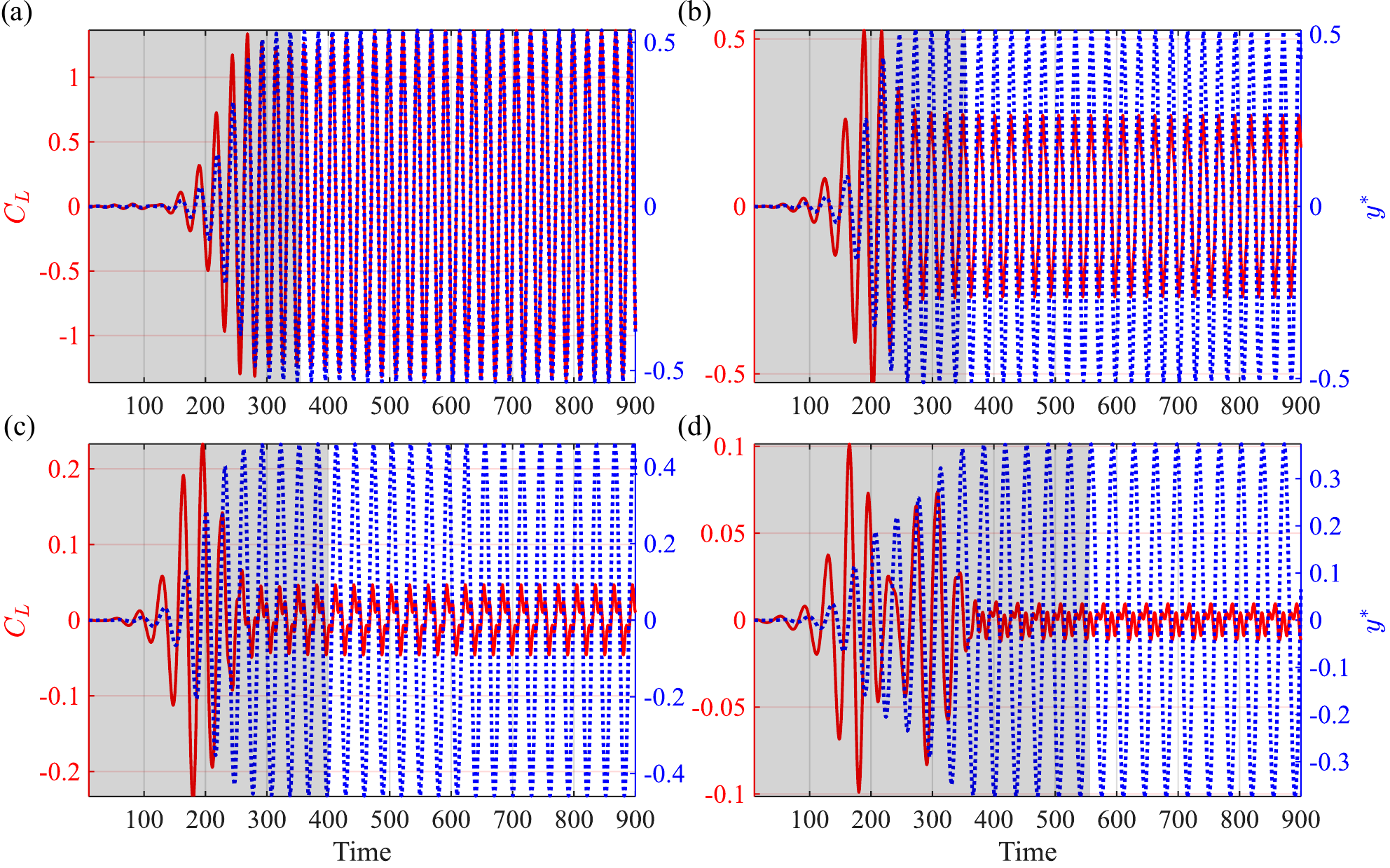}
\caption{Time histories of $C_L$ (red solid lines) and $y^*$ (blue dashed lines) in the reference FSI simulations at (a) $U^{*}=4$, (b) $U^{*}=5$, (c) $U^{*}=6$, and (d) $U^{*}=7$. The training dataset is represented by the gray shading in each case.}
\label{fig:VIV displacement}
\end{center}
\end{figure}

Figure~\ref{fig:VIV displacement} presents the time histories of $C_L$ and $y^*$ in the reference FSI simulations. The training dataset for the SINDYc employs the data during the time intervals indicated by the gray shading in Fig.~\ref{fig:VIV displacement}. In all cases, a third-order polynomial is employed for the library of candidate nonlinear functions in SINDYc. The sparsity-promoting parameter of $\alpha$ for the cases of $U^{*}=4$, $5$, $6$, and $7$ are set to be $0.001$, $0.001$, $0.001$, and $0.003$, respectively.

\begin{figure}
  \begin{center}
\includegraphics[width=95mm]{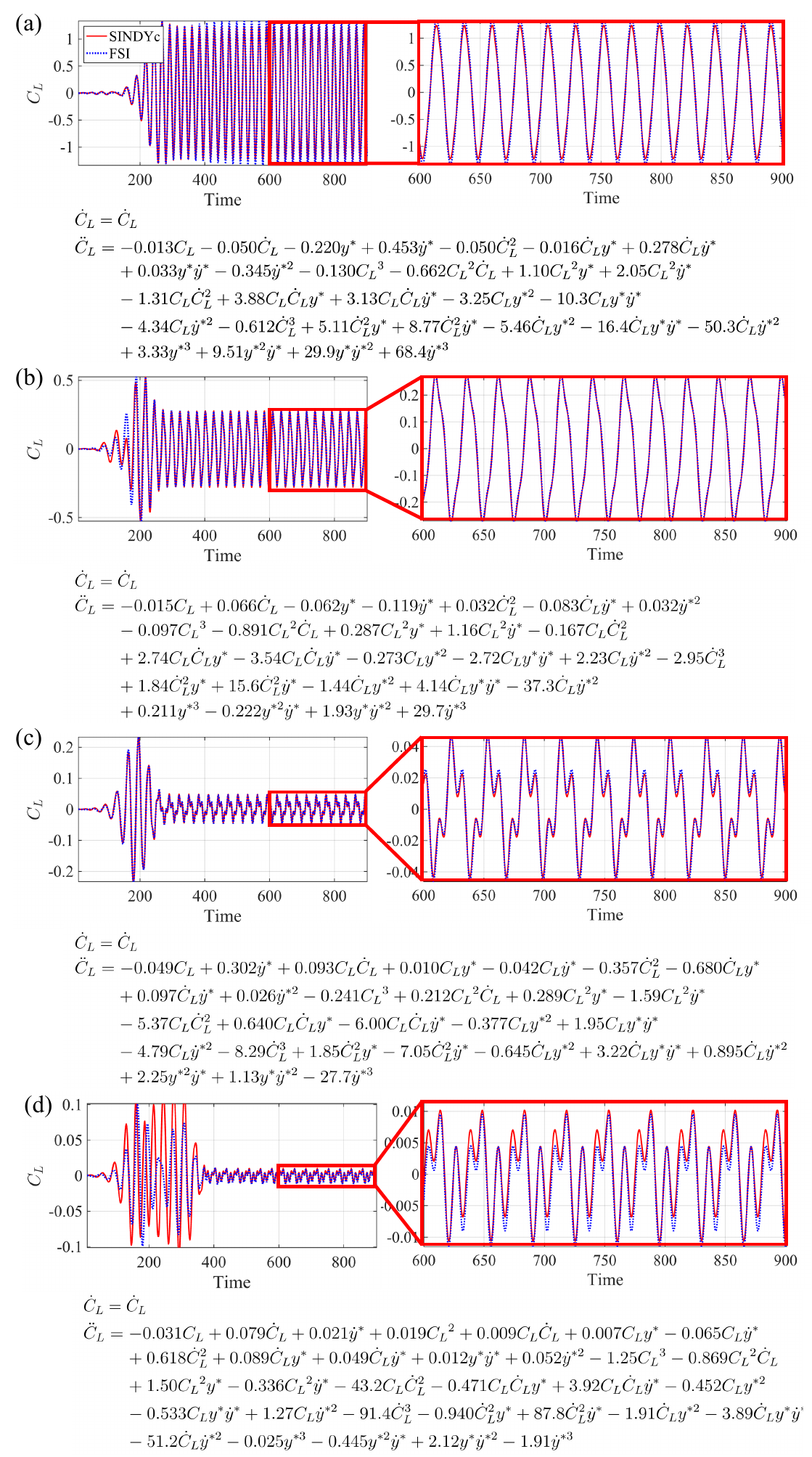}
\caption{SINDYc predictions of $C_L$ and the corresponding ordinary differential equations for (a) $U^{*}=4$, (b) $U^{*}=5$, (c) $U^{*}=6$, and (d) $U^{*}=7$. The red solid lines and blue dashed lines indicate the SINDYc and the reference FSI simulation, respectively.}
\label{fig:VIV discrete Validation}
\end{center}
\end{figure}

Figure~\ref{fig:VIV discrete Validation} represents a comparison between $C_L$ predicted by the SINDYc model and that obtained from the reference FSI simulation. The ordinary differential equations for $C_L$ corresponding to each reduced velocity $U^*$ are represented in the right column, wherein $\dot{C}_L$ was exactly recovered as $\dot{C}_L=\dot{C}_L$. Note that inputs and initial conditions for the SINDYc model were set to be those of the reference FSI simulation, i.e.,  $\textbf{U}$ and $\textbf{x}_1$. Under the conditions of $U^{*}=4-6$, SINDYc demonstrates good predictive accuracy, as shown in Fig.~\ref{fig:VIV discrete Validation}(a)-(c). For $U^{*}=7$, SINDYc also shows good agreement with the reference FSI during the periodic oscillation; however, the predictive accuracy decreases in the transient regime particularly around $t=200-300$, as presented in Fig.~\ref{fig:VIV discrete Validation}(d).
This error could be related to the complex pattern of $C_L$ behavior within the transition region. Figure~\ref{fig:VIV displacement}(d) showed that for $U^*=7$, the amplitude of $C_L$ during periodic oscillation is comparatively small, and reaching a quasi-steady state requires a longer duration. Furthermore, in the transition region of $U^*=7$ case, $C_L$ demonstrates complex behavior characterized by two amplifications at approximately $t=200$ and $300$. In other cases, $C_L$ attains only one amplification within the transition region.
Eventually, Fig.~\ref{fig:VIV discrete Validation}(d) illustrates a decrease in accuracy within the $C_L$ transition region ($t<400$); meanwhile, the accuracy in the periodic oscillation is better than that during transient regime.

Based on the prior observation, the error during the periodic oscillation is precisely quantified by comparing the reference FSI results with the predicted $C_L$, as defined by
\begin{eqnarray}
\epsilon_{C_L}:= {\sqrt{\frac{1}{N_{\text{peri}}}\sum_{i} \left(C_{L,i}^\text{FSI}-C_{L,i}^\text{SINDYc}\right)^2}}\Bigg/{\sqrt{\frac{1}{N_{\text{peri}}}\sum_{i} \left(C_{L,i}^\text{FSI}\right)^2}},\label{eq:err}
\end{eqnarray}
where $C_{L,i}^\text{FSI}$ and $C_{L,i}^\text{SINDYc}$ denote the lift coefficients at $i$-th time step obtained from the reference FSI simulation and those predicted by SINDYc, respectively, over the time intervals $t=600-900$ shown in the right panel of Fig.~\ref{fig:VIV discrete Validation}. In Eq.~\eqref{eq:err}, $\epsilon_{C_{L}}$ is defined as a normalized root-mean-square error; $N_{\text{peri}}$ denotes the number of samples, which was set as $N_{\text{peri}}=58200$. The corresponding errors for $U^{*} = 4$, $5$, $6$, and $7$ are $\epsilon_{C_{L}}=0.097$, $0.035$, $0.075$, and $0.250$, respectively.
The errors in the periodic regime are lower than 10\% for $U^*=4-6$, whereas they increase to 25\% in the case of $U^*=7.0$.
This indicates that difficulties in prediction within the transient regime affect accuracy in the periodic regime.
Improved prediction accuracy in both transient and periodic regimes may be attained through precise adjustment and identification of the SINDYc model. Meanwhile, our primary focus is the prediction of limit-cycle oscillation for VIV, and the accuracy of the current SINDYc models is considered sufficient, including the case of $U^*=7$.

\subsection{Demonstration of the proposed method}
\begin{figure}
  \begin{center}
\includegraphics[width=130mm]{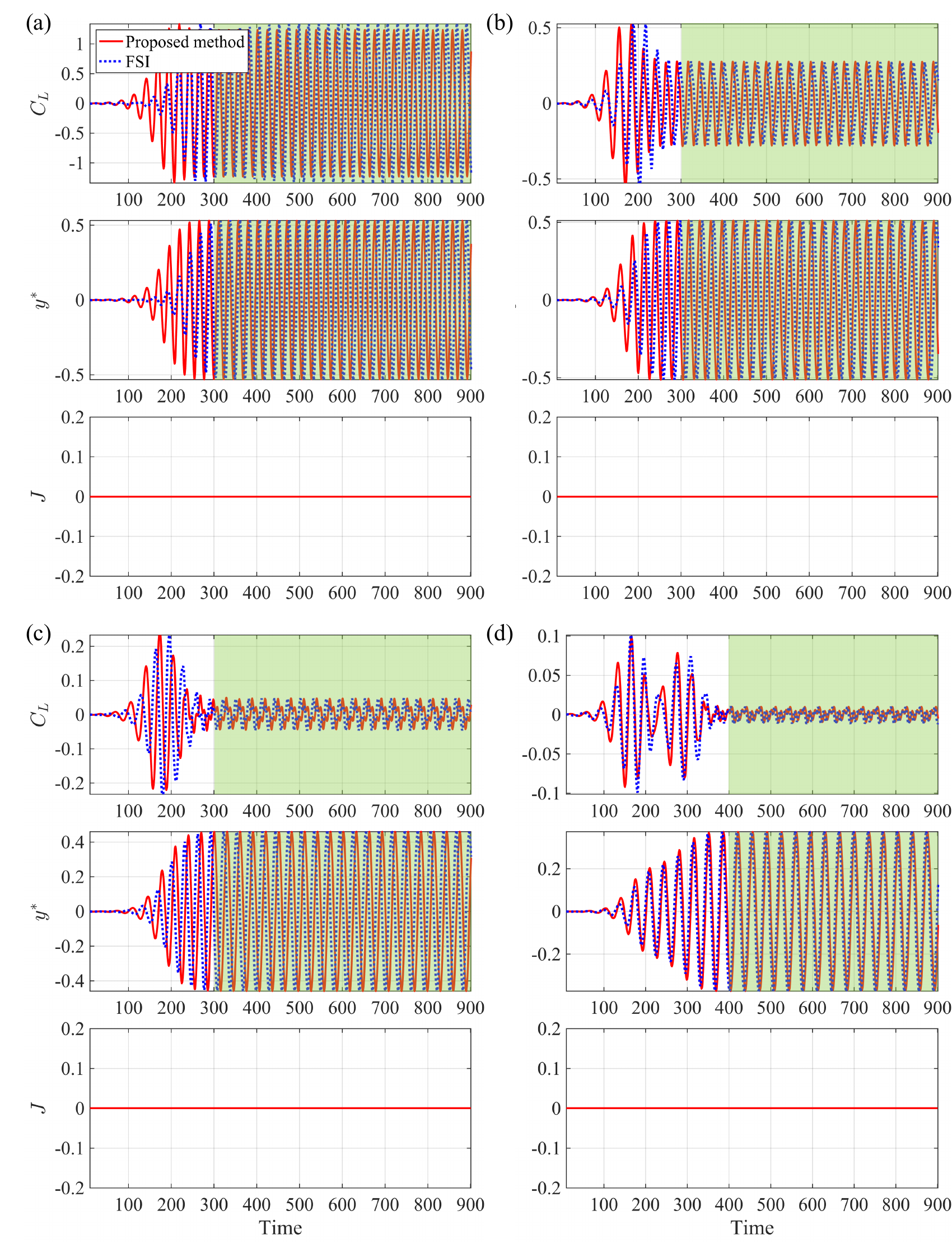}
\caption{Time histories of $C_L$ and $y^*$ obtained from the proposed method and the reference FSI simulation: (a) $U^{*}=4$, (b) $U^{*}=5$, (c) $U^{*}=6$, and (d) $U^{*}=7$. The red solid lines and blue dashed lines indicate the proposed method and reference FSI simulation, respectively. The green shading represents a limit-cycle regime in each case. The initial conditions for the proposed method are set to be the same as the reference FSI simulation.}
\label{fig:VIV without residual force}
\end{center}
\end{figure}

Two results are presented for demonstrating the proposed method. The initial result is a demonstration to predict $C_L$ and $y^*$ for the training VIV dataset, as provided by the reference FSI simulation.
In the proposed method, the initial conditions for $\textbf{x}_1$, $\textbf{u}_1$, and $\textbf{u}_2$ were set to match those of the reference FSI simulation.
The control parameter was set as $\lambda=0.5$. Figure~\ref{fig:VIV without residual force} presents the predicted physical quantities alongside the reference FSI simulation results. The residual force at the initial time step is close to zero, as the initial conditions for the proposed method were obtained from the reference FSI simulation, wherein the fluid and structural systems are already coupled. 
While the predicted $C_L$ and $y^*$ do not align with the reference FSI simulation, particularly around the transient regime around $t\simeq 200$, the periodic oscillations are well reproduced. In particular, the predicted amplitudes and oscillation patterns of $C_L$ and $y^*$ in the limit-cycle regime highlighted in green show good agreement with the reference data.
The discrete model for fluid force utilizing SINDYc does not perfectly match the reference simulation, as suggested in Fig.~\ref{fig:VIV discrete Validation}. Consequently, the prediction of the coupled state based on the SINDYc model and discrete structural system exhibited distinct behavior from the reference data, especially during the transient period. This results in a shift in the phase of periodic oscillation; nevertheless, the amplitude and frequency are accurately represented.


\begin{table}[htbp]
\caption{Initial values used in Fig.~\ref{fig:limit cycle}:
$\textbf{x}_1=(C_{L,1},\dot{C}_{L,1}),\textbf{u}_1=(y^*_1,\dot{y}^*_1)$, and $\textbf{u}_2=(y^*_2,\dot{y}^*_2)$ given as initial conditions for the reference FSI simulation and for the random case 1 through 5, consistently under different $U^*$ conditions.}
\centering
\begin{tabular}{cccc}\hline\hline
values & $\textbf{x}_1$ & $\textbf{u}_1$  & $\textbf{u}_2$  \\ \hline
reference FSI     &$(0.00,0.00)$ & $(0.00,0.00)$ &$(0.00,0.00)$    \\
random case 1         &$(0.30,0.88)$ &$(0.29,0.93)$  &$(0.54,0.65)$ \\
random case 2         &$(0.97,0.59)$ &$(0.98,0.74)$  &$(0.73,0.30)$ \\
random case 3         &$(0.60,0.74)$ &$(0.36,0.56)$  &$(0.90,0.91)$ \\
random case 4         &$(0.49,0.12)$ &$(0.60,0.49)$  &$(0.48,0.40)$ \\
random case 5         &$(0.48,0.75)$ &$(0.10,0.37)$  &$(0.23,0.18)$ \\
\hline \hline
\end{tabular}
\label{tab:init}
\end{table}

The second result demonstrates the ability to control the residual force to zero through the proposed interface control principle.
In the beginning of the simulation, the residual force $J$ was artificially generated by setting random initial conditions of $\textbf{x}_1$, $\textbf{u}_1$, and $\textbf{u}_2$.  The initial conditions were randomly selected within the range of 0.1 to 1.5, which lies below the maximum amplitudes of $C_L$ and $y^*$ observed at a reduced velocity of $U^{*} = 4$ in the reference FSI simulation. 
Five sets of initial conditions were generated randomly, designated as random case 1 through 5, and are summarized in Table.~\ref{tab:init}.
The control parameter was set as $\lambda=0.5$. 
In addition to the five specified conditions, a broader range of random values and control parameters was investigated, revealing that certain simulations exhibited instability and resulted in divergence.  The robustness of these parameters will be addressed in the next subsection.

\begin{figure}
  \begin{center}
\includegraphics[width=130mm]{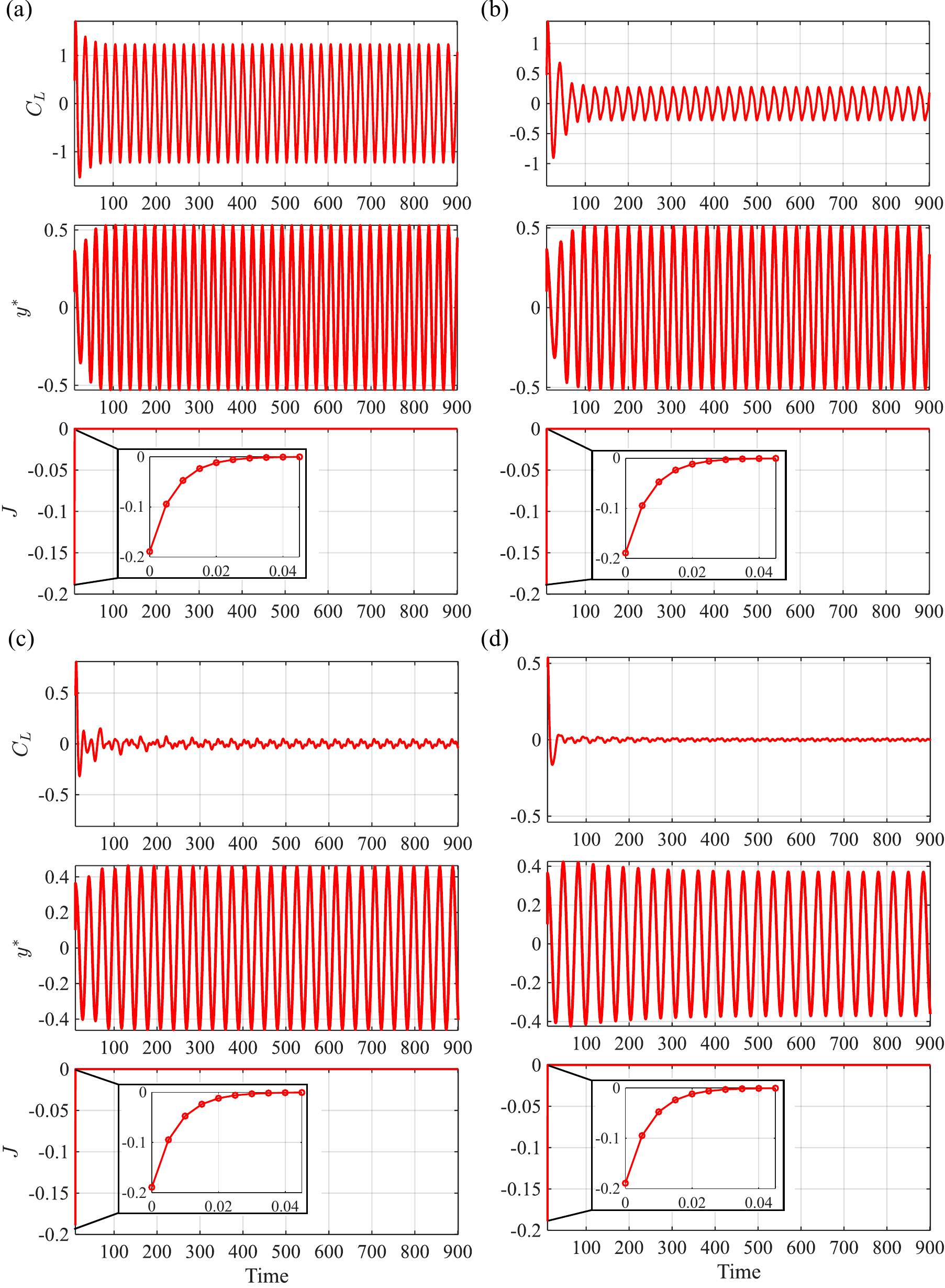}
\caption{Time histories of $C_L$, $y^*$, and $J$ predicted by the proposed method: (a) $U^{*}=4$, (b) $U^{*}=5$, (c) $U^{*}=6$, and (d) $U^{*}=7$. The initial decay behavior of the controlled residual force $J$ is enlarged and displayed in each $U^*$ case. The initial conditions for the proposed method are set to be the random case 5 in Table.~\ref{tab:init}.}
\label{fig:VIV with residual force}
\end{center}
\end{figure}

Figure~\ref{fig:VIV with residual force} presents the evolution of the controlled residual force $J$ along with $C_L$ and $y^*$ predicted by the proposed method, using the random case 5 in Table.~\ref{tab:init}. The residual force $J$ exhibits significant decreases over time in each case, indicating that the proposed control principle works effectively across all $U^*$ cases. Meanwhile, $C_L$ is relatively large during the time intervals while the residual force $J$ is present, particularly for $U^{*}=5$, $6$, and $7$. The large amplitude of $C_L$ diminishes over time, transitioning to periodic oscillations known as limit-cycle oscillations. The transient behaviors of those oscillations exhibit qualitative agreement with the reference FSI simulation, as observed in Fig.~\ref{fig:VIV displacement}. 

\begin{figure}
  \begin{center}
\includegraphics[width=135mm]{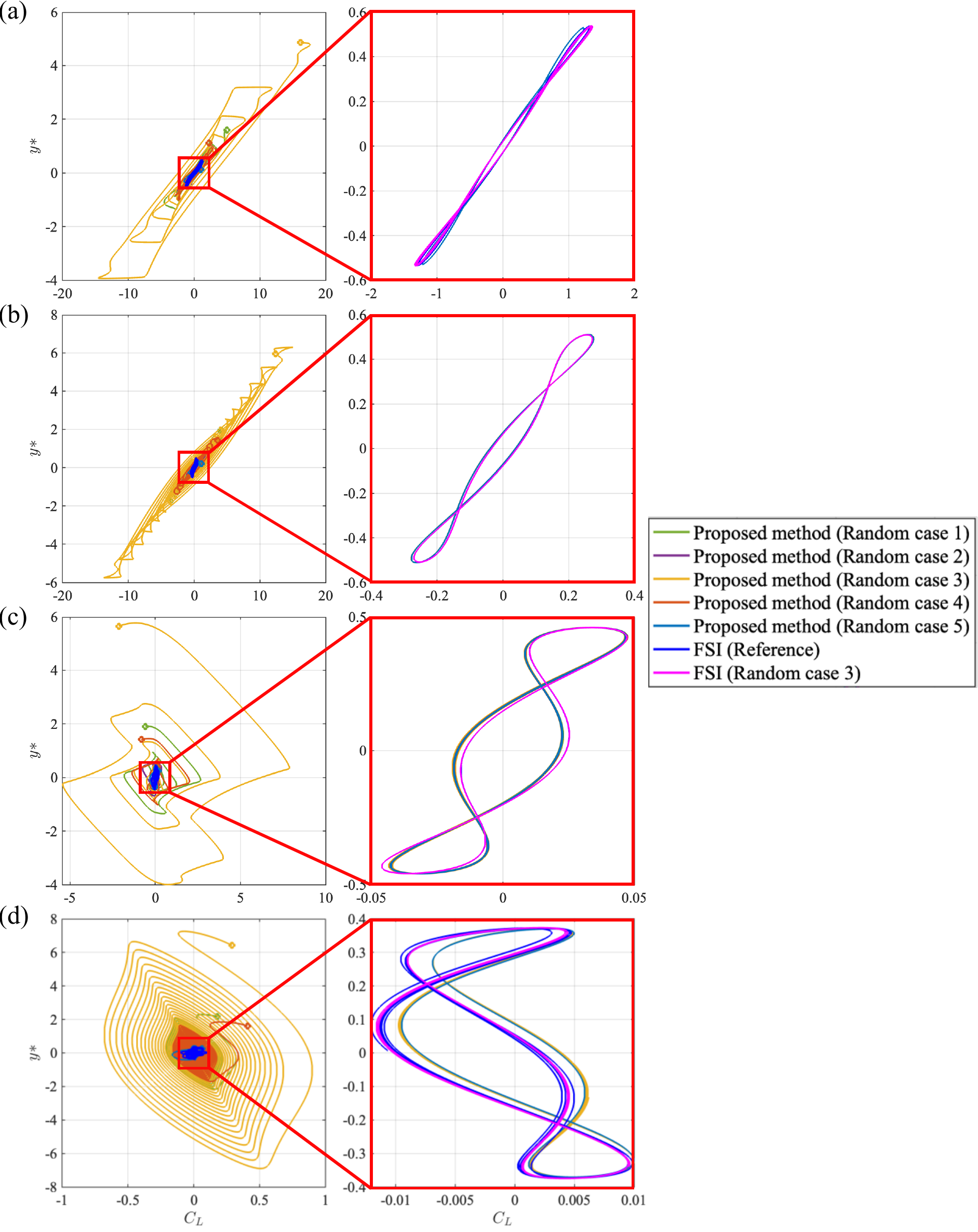}
\caption{Limit-cycle trajectories in $C_L$-$y^*$ plane are shown for the proposed method with various initial conditions of random case 1 through 5 and for the FSI simulations. Different $U^*$ cases are summarized as (a) $U^{*}=4$, (b) $U^{*}=5$, (c) $U^{*}=6$, and (d) $U^{*}=7$. The plots on the left show the full-time domain trajectory, whereas the right plots focus on the periodic oscillation period. The small circles in left images indicate the moment when the residual force $J$ first approaches zero. 
}
\label{fig:limit cycle}
\end{center}
\end{figure}

Next, a more quantitative analysis of the transient and periodic behaviors is performed for various random initial conditions listed in Table~\ref{tab:init}, in comparison with the corresponding FSI simulations.
The control parameter was set to be $\lambda=0.5$.
Figure~\ref{fig:limit cycle} shows the trajectories on $C_L$-$y^*$ space.
Additional FSI simulations were conducted for each $U^*$, wherein $y^*$ and $\dot{y}^*$, at the moment when $J$ first approaches zero in the proposed method with the random case 1 to 5, were employed as initial conditions. For brevity, only the results of the random case 3 are presented, as the other random cases exhibit similar trends. The trajectories were presented alongside those of the aforementioned reference FSI simulations.
Each panel of Fig.~\ref{fig:limit cycle} presents, on the left, the trajectory over the full time domain, and on the right, the trajectories limited to the time intervals that exhibit periodic oscillations. All the trajectories of the proposed method, although initially varying due to random initial conditions, consistently converge to distinct periodic trajectories, resulting in limit cycles.
Furthermore, the navy-blue and magenta lines, denoted as FSI (Reference) and FSI (Random case 3), represent two FSI simulations with different initial conditions described above, both converging to nearly identical periodic trajectories in each $U^*$ case.
These observations indicate that the intrinsic limit cycle of VIV is robust to variations in the initial fluid force and structural displacement.

The trajectories of the proposed method and FSI simulations exhibit an overall agreement during the periodic regime, although discrepancies are observed between them.
These discrepancies become more pronounced in the case of $U^*=7$.
This is primarily due to the error in the SINDYc predictions for the periodic regime, which was previously addressed for Fig.~\ref{fig:VIV discrete Validation}.
The proposed method effectively reproduces limit-cycle oscillations across various initial conditions for a range of $U^*$ in the VIV situation. The significant aspect of the proposed method is that it does not necessitate initial conditions in the coupled state, allowing the fluid and structural system to start from arbitrary values. The residual force remaining in the uncoupled state is subsequently reduced to zero by ensuring an appropriate displacement behavior, which can be passed to the fluid and structural systems, independently, over time.

\subsection{Robustness of the proposed method}
As mentioned in the previous subsection, the proposed method requires the specification of two types of parameters: the initial conditions ($\textbf{x}_1$, $\textbf{u}_1$, and $\textbf{u}_2$) and the control parameter ($\lambda$).  This subsection examines a broader range of parameter sets to investigate the robustness of the proposed method.
Initial conditions are generated across five specified value ranges: range 1 ($10^{-3}$-$10^{-2}$), range 2 ($10^{-2}$-$10^{-1}$), range 3 ($10^{-1}$-$10^{0}$), range 4 ($10^{0}$-$10^{1}$), and range 5 ($10^{1}$-$10^{2}$).  The random values within ranges 1, 2, and 3 are either lower than or comparable to the maximum amplitudes of $C_L$ and $y^*$ observed in the reference FSI simulation, while the values in ranges 4 and 5 exceed those amplitudes.  Fifty random initial conditions are selected within each range.  Furthermore, consistent sets of random initial conditions are utilized across all $U^*$ cases.  The control parameters are generated within the range of 0.1 to 0.9.

\begin{figure}
  \begin{center}
\includegraphics[width=135mm]{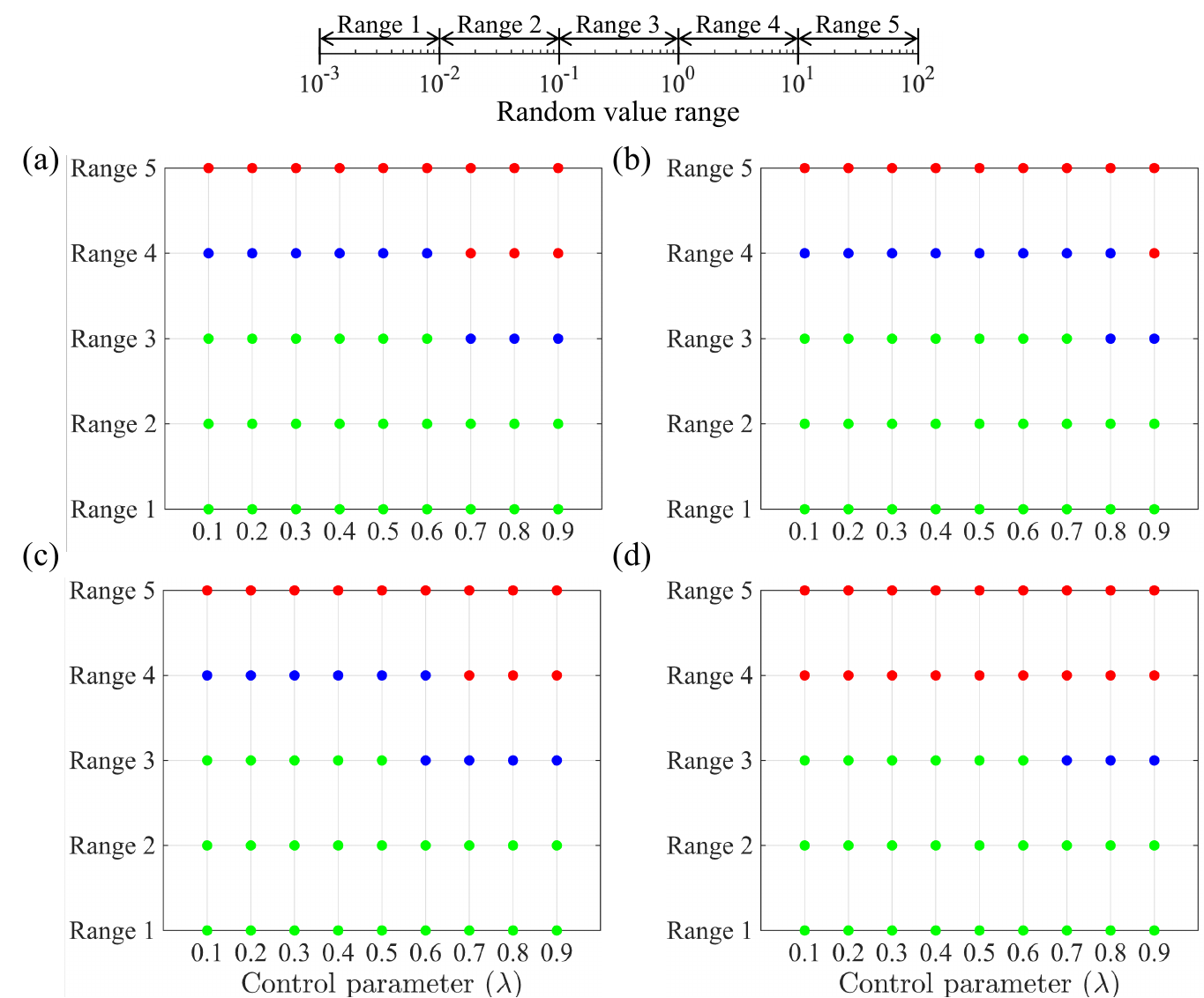}
\caption{Robustness analysis of the proposed method with various combinations of the control parameter $\lambda$ and random initial conditions: (a) $U^{*}=4$, (b) $U^{*}=5$, (c) $U^{*}=6$, and (d) $U^{*}=7$. The green and red dots show that all of the simulations are stable and unstable, respectively. The blue dots show intermediate cases where some simulations became unstable. }
\label{fig:parameter}
\end{center}
\end{figure}

Figure~\ref{fig:parameter} shows the results for examining robustness of the proposed model.
In Fig.~\ref{fig:parameter}, fifty simulations were performed for each dot, indicating the control parameter and the range of random initial conditions.
The green dots indicate that all simulations demonstrate robustness and successfully achieve periodic oscillations.  The red dots represent that all simulations exhibited instability and diverged prior to reaching the limit cycle.
The blue dots represent intermediate cases in which certain simulations became unstable and did not achieve complete stability.
All simulations are stable for the ranges 1 and 2 regardless of $U^*$.  In contrast, the robustness in range 3 varies according to the control parameter. Simulations exhibit instability within the ranges of 4 and 5, irrespective of the control parameter. The observations indicate that the proposed method demonstrates high robustness for combinations where the control parameter approaches zero and the initial conditions of $C_L$ and $y^*$ are comparable to or less than their maximum amplitudes in each reference FSI simulation.


\section{Conclusions}
In the present work, the interface control principle was introduced for a residual force minimization method to solve an unsteady FSI problem. This method enables the integration of fluid and structural domains without requiring an iterative procedure, utilizing control theory to establish an interface state that eliminates residual force over time. This study approximates a dynamical system of the interface state using reduced-order modeling, specifically the SINDYc model for fluid force and the equation of motion for structural displacement, while regulating the interface state to minimize residual force.
The VIV of a cylinder was utilized for demonstration, with reduced velocities varied for $U^{*}=4$, $5$, $6$, and $7$, which correspond to the lock-in regime.

First, the discrete fluid-force system developed by the SINDYc model was validated against the training dataset provided by the reference FSI simulation.  The accuracy was affected by the prediction capability of the transient regime for the VIV; however, the discrete accuracy of the model was adequate for predicting the limit-cycle oscillation of the cylinder displacement.
Then, the proposed method for controlling the residual force was examined in comparison to the reference FSI simulation. The results of the proposed method can vary based on the control parameters and the initial conditions of the fluid force and structural displacement. The predicted fluid force and structural displacement, initiated from the initial conditions of the reference FSI simulation, correspond closely with the oscillation patterns and amplitudes observed in the FSI simulation, regardless of the control parameter. In such a case, the residual force was not present in the initial state and was successfully kept at zero through interface control during the simulation.

Next, the demonstration focused on cases where the initial conditions deviated from those of the reference FSI simulations, aiming to demonstrate the capability to control the residual force for minimization over time and simulate limit-cycle oscillation.  The simulations investigated various random initial conditions across different reduced velocity cases of VIV. The results indicate that the proposed method effectively simulated limit-cycle oscillations subsequently to the elimination of residual forces over the initial transient phase. Meanwhile, the accuracy of the resultant amplitude and oscillation pattern during limit-cycle oscillation is dependent on the quality of the SINDYc model for the fluid force system.

Finally, the proposed method was evaluated for its robustness in achieving limit-cycle oscillation across a varied set of control parameters and initial conditions.
The results show that, across all reduced velocity cases of the VIV, the proposed method demonstrates significant robustness in cases where the control parameter approaches zero, alongside initial conditions of fluid force and structural displacement that are below their maximum amplitudes suggested in the reference FSI simulations. The lower control parameter results in a rapid convergence of the residual force, with a smaller deviation of the initial conditions from the reference FSI simulation leading to a lower residual force at the initial state. Therefore, the aforementioned observation supports the robustness of the proposed method.

\section*{Acknowledgements}
This work was supported by JST FOREST Program (JPMJFR2124 and JPMJFR222E), and JSPS KAKENHI (Grant Number JP24K01074 and JP24KF0006).

\bibliographystyle{elsarticle-harv}
\bibliography{xaerolab.bib}



\end{document}